\documentstyle[12pt]{article}
\def\reff #1{(\ref{#1})}
\def\bra #1{\langle{#1}|}
\def\ket #1{|{#1}\rangle}

\def\non{\nonumber}
\def\al{\alpha}
\def\be{\beta}
\def\ga{\gamma}
\def\Ga{\Gamma}
\def\la{\lambda}
\def\La{\Lambda}
\def\D{{\cal D}}
\def\L{{\cal L}}
\def\F{{\cal F}}
\def\de{\delta}
\def\cP{{\cal P}}
\def\l{\left\{}
\def\r{\right\}}
\def\d{\partial}
\def\om{\omega}
\def\d{\partial}

\def\beqn{\begin{eqnarray}}
\def\eeqn{\end{eqnarray}}
\def\beq{\begin{equation}}
\def\eeq{\end{equation}}
\def\d{\partial}
\def\+{{\dagger}}

\def\ds{\displaystyle}
\begin{document}

%%%%%%%%%%%%%%%%%%%%%%%%%%%%%%%%%%%%%%%%%%%%%%%%%%%%%%%%%%%%%%%%%%%

\begin{center}
{\Large\bf Interactions of Massive Integer-Spin Fields}\\
\vskip 2mm
S. M. Klishevich\footnote{E-mail address: klishevich@mx.ihep.su}
\\
{\it  Institute for High Energy Physics \\
 Protvino, Moscow Region, 142284, Russia}
%%%%%%%%%%%%%%%%%%%%%%%%%%%%%%%%%%%%%%%%%%%%%%%%%%%%%%%%%%%%%%%%%%
\end{center}
\begin{center}
\parbox{5.5in}{\footnotesize
\noindent
We review the interactions of massive fields of arbitrary integer spins
with the constant electromagnetic field and symmetrical Einstein space in
the gauge invariant framework. The problem of obtaining the gauge-invariant
Lagrangians of integer spin fields in an external field is reduced to purely
algebraic problem of finding a set of operators with certain features using
the representation of the higher-spin fields in the form of vectors in a
pseudo-Hilbert space. Such a construction is considered up to the second
order for the electromagnetic field and at linear approximation for
symmetrical Einstein space. The results obtained are valid for space-time
of arbitrary dimensionality.
}
\end{center}

\subparagraph{Introduction.}

At present only the superstring theory claims to have the consistent
description of interaction of the higher-spin fields. But interacting
strings describe the infinite set of fields and the question about
interaction of finite number of fields is still open.

In this paper we consider the interaction of an arbitrary massive spin-$s$
field with a homogeneous electromagnetic field up to the second order in
the strength and a symmetrical Einstein space at linear approximation in
curvature.

We consider the massive higher-spin fields in gauge invariant way
representing them as states on an auxiliary Fock space. We regard any
interaction as a deformation of initial free gauge algebra. The preservation
of gauge symmetry and absence of terms with higher derivatives ensure the
right number of physical degrees of freedom.

\subparagraph{Free massive field with integer spins.}
 Let us consider the Fock space generated by creation-annihilation operators
$\bar a_\mu,\ a_\mu$ and $\bar b,\ b$  which are vectors and scalars,
correspondingly, in the {$D$-dimensional} Minkowski space ${\cal M}_D$ and
form the following algebra
\begin{eqnarray}
\label{H-alg+}
\left[a_\mu,\bar a_\nu\right]&=&g_{\mu\nu}, \quad
a_\mu^{\dag} = \bar a_\mu,\qquad
\left[b,\bar b\right]=1,\quad b^{\dag}=\bar b.
\end{eqnarray}
where $g_{\mu\nu}$ is the metric tensor with signature $\|g_{\mu\nu}\|={\rm
diag}(-1,1,1,...,1)$. Since the metric is indefinite, the Fock space, which
realizes the representation of the Heisenberg algebra \reff{H-alg+}, is a
Pseudo-Hilbert space.

We shall describe the massive field of spin $s$ as the following vector
in the Fock space:
\begin{equation}
\label{FockM}
\ket{\Phi^s}=\sum\limits_{n=0}^s
\Phi_{\mu_1\dots\mu_{n}}(x)\bar b^{s-n}\prod_{i=1}^n\bar a_{\mu_i}\ket{0}.
\end{equation}
Coefficient function $\Phi_{\mu_1\dots\mu_n}(x)$ is a symmetrical tensor
of rank $n$ in space ${\cal M}_D$.

Let us introduce the following operators in our pseudo-Hilbert space
\begin{equation}
\label{L-m}
L_1=p\cdot a + mb ,\quad L_2=\frac 12\left(a\cdot a + b^2\right),
\quad L_0=p^2+m^2,\quad L_{-n}=L_n{}^{\dag}.
\end{equation}
Here $p_\mu=i\partial_\mu$ is the momentum operator that acts in the space
of the coefficient functions, while $m$ is an arbitrary parameter that has
dimensionality and sense of mass.
In non-interacting case one can consider such a transition as the
dimensional reduction ${\cal M}_{D+1}\to{{\cal M}_D\otimes S^1}$ with the
radius of sphere $R\sim1/m$ (refer also to \cite{Pashnev-1,Pashnev:MPL97}).

Operators \reff{L-m} satisfy the commutation relations:
\begin{equation}
\label{L-algebra0}
 \begin{array}{rclrcl}
 \left[L_1,L_{-2}\right]&=&L_{-1},
&\quad\left[L_1,L_2\right]&=&0,\\
 \left[L_2,L_{-2}\right]&=& N + \frac{D+1}2,
&\quad\left[L_0,L_n\right]&=&0,\\
 \left[L_1,L_{-1}\right]&=&L_0,
&\quad\left[N,L_n\right]&=&{}-nL_n,\quad n=0,\pm 1,\pm 2.
 \end{array}
\end{equation}
Here $N=\bar a\cdot a$ is a level operator that defines the spin of states.
So, for instance, for the state \reff{FockM} we have
$$
N\ket{\Phi^s} =s\ket{\Phi^s}.
$$

For state \reff{FockM} to describe the state with spin
$s$ one has to imposes the condition:
\begin{equation}
\label{Trace-L}
(L_2)^2\ket{\Phi^s}=0,
\end{equation}

Lagrangian for massive states can be written as an expectation value of a
Hermitian operator in the state \reff{FockM}
\begin{equation}
\label{Llagr}
\L_s=\bra{\Phi^s}\L(L)\ket{\Phi^s}, \quad \bra{\Phi^s}=\ket{\Phi^s}^{\dag},
\end{equation}
where
\beq
\label{L-action}
\L(L)=L_0-L_{-1}L_1-2L_{-2}L_0L_2
-L_{-2}L_{-1}L_1L_2
+\left(L_{-2}L_1L_1 + h.c.\right).
\eeq

Lagrangian \reff{Llagr} is invariant under gauge transformations
\begin{equation}
\label{Gauge-L}
\de\ket{\Phi^s}=L_{-1}\ket{\La^{s-1}}.
\end{equation}
as a consequence of the relation $\L(L) L_{-1}=(...)L_2$.
Here, the gauge state
$$
\ket{\La^{s-1}}=\sum\limits_{n=0}^{s-1}
\La_{\mu_1...\mu_n}\bar b^{s-n-1}\prod_{i=1}^n\bar a_{\mu_i}\ket{0},
$$
satisfies the condition
\begin{equation}
\label{traceless}
L_2\ket{\La}=0.
\end{equation}

For convenience, hereinafter we assume $m=1$.

%%%%%%%%%%%%%%
\subparagraph{Interactions of massive integer-spin fields.}
It is worth noting that the massive gauge higher-spin fields
are described by the first-class constraints only\footnote{The Hamiltonian
formulation for the gauge-invariant description of the massive fields with
spins 2 and 3 was considered in Ref. \cite{hamilton}} from point of view of
the Hamiltonian formulation. As is well-known the ``minimal''coupling
prescription breaks the right number of physical degrees of freedom. In the
considered here gauge manner of description this is represented as breaking
the gauge invariance and, as a consequence, breaking the algebra of the
first-class constraints. But if we can restore the gauge invariance by some
deformation of the Lagrangian and the transformation, hence we can restore
the algebra of the constraints. Of course, any local interaction of physical
fields (both consistent and inconsistent) can be rewritten in a gauge
invariant way with the help of Stueckelberg fields and therefore the
Stueckelberg gauge symmetry principle alone cannot serve as a fundamental
principle for fixing interactions. But the important point is that, in
general, such a procedure leads to gauge invariant interactions with higher
derivatives what is the reason why the number of degrees of freedom may
be changed despite that the gauge symmetry is preserved.

We introduce interactions by means of the ``minimal'' coupling prescription,
i.e. we replace usual momentum operators with covariant ones {$p_\mu
\to\cP_\mu$}. The commutator of the covariant momenta is proportional to the
strength of an external field:
$\left[\cP_\mu,\cP_\nu\right]\sim\F_{\mu\nu}$,
where $\F$ is strength of the electromagnetic field or Riemann tensor.

The $L$-operators cease to obey algebra \reff{L-algebra0} after substitution
of the usual momenta with the covariant ones in definitions \reff{L-m}.
Therefore, Lagrangian \reff{L-action} loses the invariance under
transformations \reff{Gauge-L}.

To restore algebra \reff{L-algebra0} we represent operators \reff{L-m} as
normal ordered functions of creation and annihilation operators as well as
of $\F$. The particular form of the operators $L_i$ can be defined from the
condition of recovering of commutation relations \reff{L-algebra0} by these
operators. We should note that it is enough to define the form of the
operators $L_1$ and $L_2$, since one can take following expressions:
\begin{equation}
\label{def}
L_0\stackrel{{\rm def}}{=}\left[L_1,L_{-1}\right],\quad
N\stackrel{{\rm def}}{=}\left[L_2,L_{-2}\right]-\frac{D+1}2.
\end{equation}
as definitions of the operators $L_0$ and $N$.

Since we have turned to the extended universal enveloping algebra, the
arbitrariness in the definition of operators\footnote{Such an arbitrariness
has been also presented earlier as an internal automorphism
of the Heisenberg-Weil algebra defining the Fock space \cite{Turbiner:97}.
But exactly the transformations depending on $\F$ are important for
us.} $a$ and $b$ appears. Besides, in the right-hand side of \reff{H-alg+},
we can admit the presence of the arbitrary operator functions depending on
$a$, $b$, and $\F$. Here, such a modification of the operators must
not lead to breaking the Jacob's identity. Besides, these operators must
restore the initial algebra in limit $\F\to 0$. However, one can
make sure that using the arbitrariness in the definition of creation and
annihilation operators, we can restore algebra \reff{H-alg+} for both the
e.m. and gravitational fields at corresponding orders.

We shall search for the operators $L_1$ and $L_2$ in the form of an
expansion \mbox{in $\F$}.

\subparagraph{Electromagnetic interaction of massive spin-$s$
field.}
Here we consider the interaction between massive gauge fields with
arbitrary integer spin $s$ and a constant e.m. field.

The commutator of covariant momenta defines the e.m. field strength
\begin{equation}
\left[\cP_\mu,\cP_\nu\right]=F_{\mu\nu}.
\end{equation}
For convenience we included the imaginary unit and coupling constant into
the definition of the strength tensor.

Let us consider the linear approximation and modify the $L$-operators by
appropriate way. So, the operator $L_1$ should be no higher than linear in
operator $\cP_\mu$, since the presence of its higher number changes the type
of gauge transformations and leads to appearance of terms with higher
derivatives in the Lagrangian that breaks the right number of physical
degrees of freedom. Therefore, at this order we shall search for operators
having the form
\begin{equation}
\label{L1anz}
L_1^{(1)}=\left(\bar aFa\right)h_0(\bar b,b)b + \left(\cP Fa\right)h_1(\bar
b,b)
+ \left(\bar aF\cP\right)h_2(\bar b,b)b^2.
\end{equation}
At the same time the operator $L_2$ cannot depend on the momentum operators
at all, since condition \reff{Trace-L} defines purely algebraic
constraints\footnote{Besides, presence of the derivatives in
$L_2$ also leads to appearance of higher derivatives in the
Lagrangian.} on the coefficient functions. Therefore, at this order we
choose the operator $L_2$ in the following form:
\begin{equation}
\label{L2anz}
L_2^{(1)}= \left(\bar aFa\right)h_3(\bar b,b)b^2,
\end{equation}
Here $h_i(\bar b,b)$ are normal ordered operator functions of the type
\beq\label{h_i}
h_i(\bar b,b)=\sum\limits_{n=0}^\infty H_n^i \bar b^n b^n,
\eeq
where $H^i_n$ are arbitrary real coefficients. We consider only the
real coefficients since the operators with purely imaginary coefficients do
not give any contribution to the "minimal" interaction.

Let us define the particular form of functions $h_i$ from the condition
recovering commutation relations \reff{L-algebra0} by operators
\reff{L1anz} and \reff{L2anz}. This algebra is entirely defined by
\reff{def} and by the following commutators
\beq\label{l_l}
[L_2,L_1]=0,
\qquad
[L_2,L_{-1}]=L_1,
\qquad
[L_0,L_1]=0.
\eeq
Having calculated \reff{l_l} and passing to the normal symbols of creation
and annihilation operators, we obtain a system of differential equations
for the normal symbols of operator functions $h_i$. For the normal symbols
of the operator functions we shall use the same notations. This does not
lead to the mess since we consider the operator functions as the functions
of two variables while their normal symbols as the functions of one
variable. Having solved the systems of the equations we obtain the
particular form of functions $h_i$ \cite{oscB}:
\beq\label{resh_i}
\begin{array}{ll}
h_0(x)=1-d_2,&\quad
h_1(x)=d_1\left(\frac12-x\right)e^{-2x}+d_2\left(\frac12+x\right),
\ds\\\ds
h_2(x)=d_1e^{-2x}+d_2,&\quad
h_3(x)=0.
\end{array}
\eeq
Here $d_1$ and $d_2$ are arbitrary real parameters.

So, normal symbols of operators $L_n$ has the following form in this
approximation:
\begin{eqnarray*}
\label{L^(1)}
L_1^{(1)}&=&
(1 - d_2)\left(\bar\al F\al\right)\be
+ \left(e^{-2\bar\be\be}d_1\left(\frac12 - \bar\be\be\right)
 + d_2\left(\frac12 + \bar\be\be\right)\right)(\cP F\al)
\\&&{}
 + \left(e^{ - 2 \bar\be\be} d_1 + d_2\right)\left(\bar\al F\cP\right)\be^2,
\\
L_0^{(1)}&=&
 (1 - 2 d_2)\left(\bar\al F\al\right)
 + \l (1 + 2 d_2)\left(\cP F\al\right)\bar\be + h.c.\r,
\\
L_2^{(1)}&=&0,
\end{eqnarray*}
where $\bar\al_\mu$ and $\al_\mu$ are the normal symbols of the 
operators~$\bar a_\mu$ and~$a_\mu$.

The transition to the operator functions is realized in the conventional
manner:
$$\left.
:\!O(\bar a,\bar b,a,b)\!:=
e^{\bar a\cdot\frac{\partial}{\partial\bar\al}}
e^{\bar b\frac{\partial}{\partial\bar\be}}
e^{a\cdot\frac{\partial}{\partial\al}}
e^{b\frac{\partial}{\partial \be}}
 O(\bar\al,\bar\be,\al,\be)
\right|_{{\al^\#\to0}\atop{\be^\#\to0}}.
$$

 Thus, we have obtained the general form of operators $L_n$ which obey
algebra \reff{L-algebra0} in linear approximation. This means that
Lagrangian \reff{L-action} is an invariant under
transformations~\reff{Gauge-L} at this order.

From \reff{L^(1)} it is clear that there exists the two-parametric
arbitrariness in linear approximation. But one of the arbitrary
parameters $d_1$ and $d_2$ is determined in the second approximation.
In this, there are two solutions: when $d_1$ vanishes and $d_2$ is
arbitrary, and vice versa, when $d_1$ is a free parameter and $d_2$ is
equal to $\frac12$. One can verify that the gyromagnetic ratio vanishes in
the second case.

It is worth noting that the construction obtained is free
from pathologies. Indeed, the gauge invariance and absence of higher
derivatives ensure the appropriate number of physical degree of freedom. In
this, the model is causal in linear approximation
since by virtue of the antisymmetry and homogeneity of $F_{\mu\nu}$ the
characteristic determinant\footnote{The determinant is entirely determined
by the coefficients of the highest derivatives in equations of motion after
gauge fixing and resolving of all the constraints \cite{Velo1:69}.}
for equations of motion of any massive state has the form $D(n)=(n^2)^p +
{\cal O}\left(F^2\right)$, where $n_\mu$ is a normal vector to the
characteristic
surface and the integer constant $p$ depends on the spin of massive state.
The equations of motion will be causal (hyperbolic) if the solutions $n^0$
to $D(n)=0$ are real for any $\vec n$. In our case condition $D(n)=0$
corresponds to the ordinary light cone at this order.

One can consider the next approximation in the similar manner \cite{oscB}.
The only essential difference of this order is that the operator $L_2$
depends on $F_{\mu\nu}$.

Thereby, we have restored algebra \reff{L-algebra0} up to the second order
in the electromagnetic field strength. It means that we have restored the
gauge invariance\footnote{Which ensures the right number of physical degree
of freedom since in this order there are no terms with higher derivatives
either.} of Lagrangian \reff{L-action} at the same order as well. Here
we have not used the dimensionality of space-time anywhere explicitly, i.e.
the obtained expressions do not depend on it.

\subparagraph{Propagation of massive higher-spin field in symmetrical
Einstein space.} 
Now we consider an arbitrary $D$-dimensional symmetrical
Einstein space, i.e. the Riemann space defined by the following equations:
$$
R_{\mu\nu} -\frac12g_{\mu\nu}R + g_{\mu\nu}\la=0,
\qquad
\D^{\left(\Ga\right)}_\mu R_{\nu\al\be\ga} = 0,
$$
where $\D^{(\Ga)}_\mu$ is the covariant derivative with the Cristofel
connection
$\Ga^\al{}_{\nu\mu}$. We assume that the Greek indexes are global while the
Latin ones are local. As usual, the derivative $\D^{(\Ga)}_\mu$ acts on
tensor fields with global indexes only.

To describe the massive higher-spin fields in the Riemann background, we
must replace the ordinary derivatives with the covariant one, i.e. we make
the substitution:
\beq\label{covariant}
p_\mu\to\cP_\mu=i\left(\D^{(\Ga)}_\mu+\omega_\mu{}^{ab}\bar a_aa_b\right),
\eeq
where $\om_\mu^{ab}$ is the Lorentz connection. We imply that the creation
and annihilation operators primordially carry the local indexes. We also
have to introduce the non-degenerate vielbein $e^a_\mu$ for the transition
from the local indexes to the global ones and vice versa. As usual, we
impose the conventional requirement on the vielbein
$$
\D_\mu^{\left(\Ga+\om\right)} e^a_\nu =
\d_\mu e_\nu^a - \Ga^\la{}_{\nu\mu}e_\la^a + \om_\mu{}^a{}_be^b_\nu=0.
$$
By means of this relation one can transfer from expressions with one
connection to those with other. Besides, we should note that due to this
relation the operator $\cP_\mu$ commutes with the vector
creation-annihilation operators carrying global indexes
$\bar a_\nu=e_\nu^b\,\bar a_b$ and $a_\nu=e_\nu^b\,a_b$. This allows us not
to care about the ordering of operators \reff{L-m}.

 One can verify that the covariant momentum operator defined in this way
 properly acts on the states of type \reff{FockM}, indeed
$$
\cP_\mu\ket{\Phi}=i\D_\mu^{\left(\om\right)}\Phi^{b_1\dots b_n}
\prod_{i=1}^n\bar a_{b_i}\ket{0}
 = i\D_\mu^{\left(\Ga\right)}\Phi^{\nu_1\dots\nu_n}\prod_{i=1}^n\bar
a_{\nu_i}\ket{0}.
$$

 The commutator of the covariant momenta defines the Riemann tensor:
\begin{equation}
\left[\cP_\mu,\cP_\nu\right]=R_{\mu\nu}{}^{ab}\left(\om\right)\bar a_aa_b.
\end{equation}
where $R_{\mu\nu}{}^{ab}\left(\om\right)=\d_\mu\om_\nu{}^{ab}
+\om_\mu{}^a{}_c\,\om_\nu{}^{cb} - \left(\mu\leftrightarrow\nu\right)$.

In the definition of operators \reff{L-m}, we replace the ordinary momenta
with the covariant ones as well. As a result, the operators cease to obey
algebra~\reff{L-algebra0}. Therefore, Lagrangian \reff{L-action} loses the
invariance under gauge transformations~\reff{Gauge-L}.

To recover the gauge invariance, we do not need to restore the whole
algebra~\reff{L-algebra0}, it is enough to ensures the existence of the
two first commutators in \reff{l_l}. To restore these relations, we shall
search for the operators $L_1$ and $L_2$ as series in the Riemann
tensor and scalar curvature.

Let us consider linear approximation.

Using the very arguments as for the case with constant e.m. field we choose
the following ansatz
\beqn\label{L1anzR}
L_1^{(1)}&=&
	 R\biggl(h_0(\bar b,b)\,b
 + h_1(\bar b,b)\,b\left( \bar a\cdot a\right)
 +  \bar b h_2(\bar b,b)\,a^2 + h_3(\bar b,b)\,b^3\bar a^2
\non\\&&{}
 + h_4(\bar b,b)\left(\cP\cdot a\right) + h_5(\bar b,b)\,b^2\left(\bar a
\cdot\cP\right)
\biggr)
	 +  R^{\mu\nu ab}\biggl(
	   h_6(\bar b,b)\, b\bar a_\mu\bar a_a a_\nu a_b
\non\\&&{}
	 + h_7(\bar b,b)\,\bar a_\mu a_\nu \cP_a a_b
	 + h_8(\bar b,b)\, b^2\bar a_\mu\cP_\nu\bar a_a a_b\biggr).
\eeqn
for operator $L_1$ at this approximation and
\beqn\label{L2anzR}
L_2^{(1)}&=&
          R\left( h_9(\bar b,b)\, b^2
	+ h_{10}(\bar b,b)\,a^2
        + h_{11}(\bar b,b) b^2\left(\bar a\cdot a\right)
        + h_{12}(\bar b,b)\, b^4\bar a^2\right)
\non\\&&{}
	+ h_{13}(\bar b,b)\,b^2 \bar a^\mu\bar a^a a^\nu a^b R_{\mu\nu ab}.
\eeqn
Here $h_i(\bar b,b)$ are normal ordered operator functions of type
\reff{h_i}.

Let us define a particular form of the functions $h_i$ from the condition
of recovering two first commutation relations in \reff{l_l} by the
operators~$L_1$ and~$L_2$.

We have to note that these operators can obey the two first relations in
\reff{l_l} up to the terms proportional to $L_2^{(0)}=\frac12\left(a^2+b^2
\right)$ at
linear order, since this does not break the gauge invariance due to
constraint~\reff{traceless}.

Having calculated the commutators under consideration and passing to normal
symbols of the creation and annihilation operators, we obtain a system of
differential equations in the normal symbols of operator functions $h_i$.
Having solved that, we obtain the particular form of the operators $L_1$
and~$L_2$ \cite{HSRiemann}:
\beqn\label{Resh_i}
L_1^{(1)}&=&
\frac{1}{6}R^{\mu\nu\al\be}\bar\al_\al\al_\mu
\left\{\cP_\nu\al_\be\left(1+2\bar\be\be\right) - 5\bar\al_\nu\al_\be \be
 + 2\bar\al_\nu\cP_\be \be^2\right\}
\non\\&&{}
 + R\left\{
 c_2\left(\cP\cdot\al\right) + c_1\be
 \right\},
\non\\
L_2^{(1)}&=&
 \frac{R}{12D}\left\{\al^2\left(1 + 2\bar\be\be\right) + 4\bar\be\be^3\right
\},
\eeqn
where $c_1$ and $c_2$ are arbitrary real parameters.

 Thus, we have obtained the general form of the operators $L_n$, which
provide the gauge invariance of Lagrangian \reff{L-action} at this order.
The form of the operator $L_2$ has changed in this approximation, hence, the
conditions
$$
\left(L_2\right)^2\ket{\Phi^s}=0, \qquad L_2\ket{\La^{s-1}}=0
$$
undergo the nontrivial modifications in terms of the coefficient functions.

%%%%%%%%%%%%%%%%%%%%%%%%%
\subparagraph{Conclusion.}
In this paper we have constructed the Lagrangian describing the
interaction of massive fields of arbitrary integer spins with the
homogeneous electromagnetic field up to the second order in the strength.
It is noteworthy that unlike the string approach \cite{Argyres} our
consideration does not depend on the space-time dimensionality, and,
moreover, we have described the interaction of the single field with
spin $s$ while in the string approach the presence of constant
electromagnetic field leads to the mixing of states with different
spins \cite{spin3_1} and one cannot consider any states separately.

Consistency of the obtained results is based on the two following
principles: i) the absence of terms with higher-derivatives and
ii) gauge symmetry. Item i) provides preservation of dimensionality of the
unreduced phase space when an interaction is switched on, while
item ii) in addition to i) guarantees preservation of dimensionality of the
physical phase space. So, these requirements provide the right number of
physical degrees of freedom in the interacting case.

%%%%%%%%%%%%%%%%%%%%%%%

%%%%%%%%%%%%%%%%%%%%%%%%%%%%%%%%%%%%%%%%

\end{document}